\documentclass[preprint2]{aastex}
\newcommand{\Msun}{\ifmmode\mbox{M}_{\odot}\else$\mbox{M}_{\odot}$\fi}
\newcommand{\Rsun}{\ifmmode\mbox{R}_{\odot}\else$\mbox{R}_{\odot}$\fi}
\newcommand{\erg}{\ifmmode\textrm{erg}\else$\textrm{erg}$\fi}
\newcommand{\orb}{\mathrm{orb}}
\newcommand{\obs}{\mathrm{obs}}

\newcommand{\trial}{\mathrm{trial}}

\shorttitle{No X-ray pulsations in LMXB 4U~1820$-$30}
\shortauthors{Dib et al.}
\begin{document}
%% --------------------------------------------------------
\title{An RXTE Archival Search for Coherent X-ray Pulsations in LMXB
4U~1820$-$30}
%% --------------------------------------------------------
\author{Rim~Dib\altaffilmark{1}, Scott~M.~Ransom\altaffilmark{1},
        Paul~S.~Ray\altaffilmark{2}, Victoria~M.~Kaspi\altaffilmark{1}, and
	Andrew~M.~Archibald\altaffilmark{3}}
%% --------------------------------------------------------
%% \email{}
%% --------------------------------------------------------
\altaffiltext{1}{Department of Physics, McGill University,
                 Montreal, QC H3A~2T8}
\altaffiltext{2}{E. O. Hulburt Center for Space
                 Research, Code 7655, Naval Research Laboratory, Washington,
                 DC 20375}
\altaffiltext{3}{Department of Mathematics and Statistics , McGill 
                 University, Montreal, QC H3A 2K6}
%% --------------------------------------------------------
\begin{abstract}
%% --------------------------------------------------------
As part of a large-scale search for coherent pulsations from LMXBs in the
{\em RXTE} archive, we have completed a detailed series of searches for
coherent pulsations of 4U~1820$-$30 --- an ultracompact LMXB with a binary
period of 11.4 min, located in the globular cluster NGC~6624.  The short
binary period implies any coherent signal would be highly accelerated, so we
used phase modulation searches, orbital-parameter-fitting coherent searches,
and standard acceleration searches to give significant sensitivity to
millisecond pulsations. We searched, in four energy bands and at a range of
luminosities, a total of 34 archival {\em RXTE} observations, 32 of which
had on-source integration times longer than 10\,ks, and some of which were
made consecutively which allowed us to combine them. We found no pulsations.
Using our phase modulation search technique, which we ran on all 34
observations, we have been able to place the first stringent (95\%
confidence) pulsed fraction limits of $\lesssim$~0.8\% for all realistic
spin frequencies (i.e. $\lesssim $~2\,kHz) and likely companion masses
(0.02\,\Msun~$\leq M_c \leq$~0.3\,\Msun). Using our
orbital-parameter-fitting coherent search, which we ran on only 11 selected
observations, we have placed a pulsed fraction limit of $\lesssim$~0.3\% for
spin frequencies $\lesssim$~1.25\,kHz and companion masses $M_c
\leq$~0.106\,\Msun. By contrast, all five LMXBs known to emit coherent
pulsations have intrinsic pulsed fractions in the range 3\% to 7\% when
pulsations are observed. Hence, our searches rule out pulsations with
significantly lower pulsed fractions than those already observed.
%% --------------------------------------------------------
\end{abstract}
%% --------------------------------------------------------
%% --------------------------------------------------------
%% A small citation guide:
%% \citep{jon90} (Jones et al. 1990) [no comma] 
%% \cite(t){jon90} Jones et al. (1990) [no comma]
%%
%% \citep*{jon90} (Jones, Baker, and Williams 1990) [one arg, no comma]
%% \citet*{jon90} Jones, Baker, and Williams (1990) [one arg]
%%
%% \citealp(jon90) Jones et al. 1990  [no comma]
%% \citealt{jon90} Jones et al. 1990  [no comma, been using this]
%%
%% \citealp*{jon90} Jones, Baker, and Williams 1990 [all and no comma,this]
%% \citealt*{jon90} Jones, Baker, and Williams 1990 [all and no comma]
%% --------------------------------------------------------
%% --------------------------------------------------------
%% ----------------%#################%---------------------
%% ----------------%SECTION 1 : INTRO%---------------------
%% ----------------%#################%---------------------
%% --------------------------------------------------------
\section{Introduction}
%% --------------------------------------------------------
One of the great scientific expectations when the Rossi X-ray Timing
Explorer ({\em RXTE}) was launched in 1995 was the discovery of coherent
pulsations from low-mass X-ray binaries (LMXBs). At present, only five
accreting millisecond pulsars in LMXBs are known. All five are faint
transient sources for which the pulsations at the spin period of the pulsar
were discovered during outbursts (\citealt{source1}; \citealt{source2};
\citealt{source3}; \citealp*{source4}; \citealt{source5}).
%% --------------------------------------------------------
\par
%% --------------------------------------------------------
Surprisingly, no pulsations have ever been seen in the non-transient X-ray
emission from any LMXB. One possibility is that MSP coherent pulsations are
smeared out by rapidly changing orbital Doppler shifts in short period
binaries. Modern accelerated signal search techniques have not been applied
systematically to many LMXBs so it is possible that there are millisecond
pulsations in them that have gone unnoticed.
%% --------------------------------------------------------
\par
%% --------------------------------------------------------
There are many things that we can learn from searching for more examples of
direct pulsations from LMXBs. Coherent pulsations give the precise rotation
rate of the neutron star and test the connection between recycled MSPs and
LMXBs. They help constrain the models of kHz QPOs and burst oscillations.
Timing of the coherent pulses would also allow measurements of the accretion
torques, the orbital parameters, and potentially the companion masses.  In
addition, in combination with X-ray and optical spectra, pulse timing can
give information about the compactness and the equation of state of the
neutron star. Finally, setting stringent upper limits on coherent pulsations
in several sources in a variety of spectral states will impose constraints
on the possible mechanism for suppression of the predicted coherent
pulsations mentioned above. For these reasons, we have started a large-scale
archival {\em{RXTE}} search for coherent pulsations from LMXBs, the first
target of which is the ultracompact globular cluster LMXB 4U~1820$-$30.
%% --------------------------------------------------------
\par
%% --------------------------------------------------------
This paper is organized as follows. In Section~\ref{sec:characteristics}, we
describe the general properties of 4U~1820$-$30. In
Section~\ref{sec:observations}, we describe the observations as well as the
procedure we used to search for coherent pulsations. In
Section~\ref{sec:analysis}, we describe our search techniques and the
methods used to investigate the obtained candidate pulsation frequencies. In
Section~\ref{sec:results}, we describe our results and derive our upper
limits on possible coherent pulsations. Finally, in
Sections~\ref{sec:discussion} and~\ref{sec:sum}, we discuss the implications
of our findings and summarize.
%% --------------------------------------------------------
%% -------------%#############################%------------
%% -------------%SECTION 1.1 : CHARACTERISTICS%------------
%% -------------%#############################%------------
%% --------------------------------------------------------
\subsection{Characteristics of 4U~1820$-$30}
\label{sec:characteristics}
%% --------------------------------------------------------
The source 4U~1820$-$30 is an atoll LMXB in globular cluster NGC 6624
\citep*{spw87}. It has an orbital binary period of 685~s (11.4~min). The
orbital period was identified to be the period of a nearly sinusoidal X-ray
luminosity modulation with an amplitude of 2\% to 3\% and no strong energy
dependence \citep{spw87,smm86,sw89,tlp91,vhd+93,vhv+93,cg01}.  A 16\%
modulation at a period of 687$\pm$2.4~s was also seen in the UV~band
\citep{amd+97}. This is the shortest known binary orbital period in an LMXB.
4U~1820$-$30 undergoes a regular $\sim$176~day accretion cycle \citep{pt84}
switching between high and low luminosity states, which differ by a factor
of $\sim$3 in their luminosity \citep{spw87}. This luminosity variation
makes 4U~1820$-$30 an attractive target for our search because it allows us
to sample a fairly large range of accretion rates in case the pulsations
become visible only in some. In the low luminosity state, regular Type~I
bursts are seen $\pm$23 days around the minimum luminosity \citep{cg01}.  It
is notable that burst oscillations have not been observed from this source.
In the low state, 4U~1820$-$30 has also shown an extremely energetic
superburst \citep{sb02}. This bursting behaviour confirms that the observed
luminosity variations actually correspond to changes in the local accretion
rate at the surface, and not just a geometrical effect, since they affect
the bursting properties of the source. 4U~1820$-$30 also exhibits a variety
of spectral states as can be seen from Figure~\ref{figure1}. Several low
frequency QPOs (\citealp*{spw87}; \citealt{dmm+89}; \citealp*{wvr99}). as
well as two peaks of kHz~QPOs (\citealp*{szw97}; \citealt{zss+98};
\citealt{kpb+99}) have been observed from this source. The separation
between the two peaks is 275$\pm$8~Hz \citep{QPOreview}. If the secondary
star in the system is a white dwarf, its mass is estimated to be 0.058 to
0.078~\Msun~\citep{rmj+87}. If the secondary is a main sequence star, the
upper limit on its mass is 0.3~\Msun~\citep{ak93}. 
%% --------------------------------------------------------
\par 
%% --------------------------------------------------------
Of all known LMXBs, we searched 4U~1820$-$30 first for several reasons.
First, many high-time-resolution, long data sets are available in the
archives. Second, the presence of kHz QPOs in this source may indicate
favorable conditions for detecting pulsations. Third, its extremely short
orbital period allows the use of the phase modulation search technique. This
technique is most sensitive and provides the biggest increase in sensitivity
over previous methods when the observations are longer than two complete
binary orbits \citep*{rce03x}.  The precisely known and short orbital period
also makes 4U~1820$-$30 one of only two LMXBs that can be searched using our
fully coherent orbital-parameter-fitting search technique on a reasonable
time scale. 
%% --------------------------------------------------------
\par 
%% --------------------------------------------------------
The source 4U~1820$-$30 has previously been searched for coherent pulsations
at frequencies under 50~Hz by \cite{whn+91x} and by \cite{vvw+94x}. Their
upper limit on the pulsed fraction for sinusoidal pulses was 0.6\% (95\%
confidence level), and 1.1\% (99\% confidence level) respectively. Our
searches cover a much wider frequency range, including, for the first time,
the kHz range. This is particularly important since likely spin frequencies
are near 275~Hz or 550~Hz, based on the kHz~QPOs \citep{QPOreview,
THEnaturepaper}.
%% --------------------------------------------------------
%% -----------------%#######################%--------------
%% -----------------%SECTION 2 : PREPARATION%--------------
%% -----------------%#######################%--------------
%% --------------------------------------------------------
\section{Observations and Preliminary Data Preparation}
\label{sec:observations}
%% --------------------------------------------------------
We searched 34 archival {\em{RXTE}} PCA observations collected between 1996
and 2002 (see Table~\ref{table1}). All the observations were available in
event modes with time resolutions of $\sim$125~$\mu$s. 32 of the
observations had a total time on source $>$~10~ks.  The longest of these was
29~ks with a total time on source of 16~ks.  Some of the observations were
segments of longer ones which allowed us to concatenate and analyse them
together.  The longest of the concatenated observations was 77.5~ks with a
total time on source of 46.5~ks.
%% --------------------------------------------------------
\par
%% --------------------------------------------------------
For each observation, we downloaded the files of raw events (one file for
each {\em{RXTE}} orbit). The duration of an {\em{RXTE}} orbit (including
Earth occultation time) is $\sim$5.7~ks. We filtered the files by selecting
only events where the source elevation was $>$~10$^{\circ}$, the pointing
offset was $<$~0.02$^{\circ}$, and where a minimum number of operational
PCUs were on (see Table~\ref{table1}). We then transformed the arrival times
of the events to the solar system barycenter using the
`FTOOL'\footnote{http://heasarc.gsfc.nasa.gov/ftools/} \texttt{FAXBARY}. The
barycentering was necessary to remove the effects of the spacecraft and
Earth motions. We then split the data into four energy bands: Soft (2-5
keV), Medium (5-10 keV), Hard (10-20 keV), and Wide (2-20 keV), in order to
attempt to maximize the signal-to-noise ratio of the unknown pulsations.
This processing used custom Python scripts to call the standard `FTOOLs'. At
the end of the preliminary data preparation we obtained four lists of events
(photon arrival times) for every {\em{RXTE}} orbit in every observation,
each list corresponding to one of the four energy bands.
%% --------------------------------------------------------
%% -----------------%#########################%------------
%% -----------------%SECTION 3 : DATA ANALYSIS%------------
%% -----------------%#########################%------------
%% --------------------------------------------------------
\section{Data Analysis} 
\label{sec:analysis}
%% --------------------------------------------------------
After we prepared the {\em{RXTE}}-orbit-long lists of events, we
concatenated them into observation-long lists of events, and also broke them
into much shorter lists of events. We then binned the events into time
series of different time resolutions (see Table~\ref{table2}) and searched
them for pulsations using three types of searches which we summarize in
Sections~ \ref{sec:tech1}, \ref{sec:tech2}, and \ref{sec:tech3}. In
Sections~ \ref{sec:cand1}, \ref{sec:cand2}, and \ref{sec:cand3}, we explain
how we used the three searches to obtain candidate pulsation frequencies.
Finally, in Sections~\ref{sec:fin1}, \ref{sec:fin2}, and \ref{sec:fin3}, we
explain the methods we used to investigate the obtained lists of candidate
pulsation frequencies. All the data sets were processed on the 52-node dual
1.4~GHz Athlon-based Beowulf cluster operated by the McGill University
Pulsar Group.
%% --------------------------------------------------------
%% -----------------%##########################%-----------
%% -----------------%SECTION 3.1 : ACCEL SEARCH%-----------
%% -----------------%##########################%-----------
%% --------------------------------------------------------
\subsection{Acceleration Searches}
%% --------------------------------------------------------
\subsubsection{The Search Technique}
\label{sec:tech1}
%% --------------------------------------------------------
This is the method traditionally used to compensate for the effects of
orbital motion \citep[e.g.,][]{jk91}: a time series is stretched or
compressed appropriately to account for a trial {\em{constant}} frequency
derivative (i.e.$\,${\em{constant}} acceleration), Fourier transformed, and
then searched for pulsations. We used a Fourier-domain variant of this
technique based on matched filtering techniques which includes Fourier
interpolation and harmonic summing of 1, 2, 4, and 8 harmonics to improve
the sensitivity to non-sinusoidal pulsations \citep*{rem02x}.
%% --------------------------------------------------------
\par 
%% --------------------------------------------------------
Acceleration searches work best for systems where the orbital
period~$\simeq$~10$\times$the observation length. Acceleration searches are
not optimal for systems like 4U~1820$-$30 where the orbital period is
shorter than the observation duration: the acceleration is not constant over
the length of our time series.  However, we chose not to run searches on
time series shorter than 600~s (see Table~\ref{table2}) because we required
a large number of photons. So while we did not expect acceleration searches
to find pulsations, we still used them because they are included in our
standard search pipeline. Also, the segmented acceleration searches that we
ran, even though they were not short enough for the acceleration to be
constant, can still be useful if a strong pulsed signal appears for a short
period of time perhaps because the `weather' on the surface briefly cleared
and the pulsations appeared.
%% --------------------------------------------------------
\subsubsection{Obtaining Candidate Pulsation Frequencies}
\label{sec:cand1}
%% --------------------------------------------------------
To run acceleration searches, we first binned the lists of events which we
obtained in the data preparation into time series of different lengths and
time resolutions (see Table~\ref{table2}). We performed an FFT on each of
the time series and fed the result into a local red noise reduction
algorithm which divides the FFT into continuous intervals the size of which 
increases logarithmically with frequency. The power in every bin of every
interval is divided by the mean power in that interval. This produces FFTs
that have a reduced signature of red noise.  We then ran the acceleration
searches on the obtained FFTs, searching for a maximum constant acceleration
of $a_{\max}~<~{5.1 \times 10^{10}}/(f{T^2}_{\obs})$~m/s$^2$, where $f$ is
the frequency of the detected pulsations in Hz, and $T_{\obs}$ is the
integration time in seconds.  For a signal of period 3~ms and an integration
time of 600~s, this acceleration is 425~m/s$^2$. This number was chosen
because it corresponds to the maximum length of a trial Fourier response
template used in the search that would fit in the cache memory of the
available machines. The maximum possible line of sight acceleration of
4U~1820$-$30 is $a_{\max}(x) = 25240.6\times x$~m/s$^2$ where $x$ is the
projected semi-major axis of the binary in light-seconds. For a companion
mass of 0.05~\Msun$\;$ and an inclination angle of 60$^{\circ}$,
$x$~=~13.4~lt-ms, and $a_{\max}$~=~339~m/s$^2$.
%% --------------------------------------------------------
\par
%% --------------------------------------------------------
Using acceleration searches, we searched for signals with an equivalent
Gaussian significance $\sigma>2$ \citep{gro75} and with frequencies between
0.0375~Hz and 2048~Hz.  We estimated $\sigma$ based on the probability
distribution of sums of powers in a normalized power spectrum
\citep{gro75,rem02x}. In order to correctly determine the significance of a
candidate, we also took into account the number of independent trials in the
searches \citep{rem02x}. Sometimes we increased the lower limit on the
searched frequency to 1/8~Hz in order to eliminate the large number of
low-frequency candidate oscillations that the red noise reduction routine
did not successfully eliminate.
%% --------------------------------------------------------
\subsubsection{Candidate Follow-up}
\label{sec:fin1}
%% --------------------------------------------------------
When our acceleration searches detect a candidate oscillation, they return
values for the frequency of the signal $f$, the frequency derivative
$\dot{f}$, and the significance of the detection $\sigma$.  Using the
returned values for $f$ and $\dot{f}$, we folded all acceleration candidates
with $\sigma>3$. Our folding software had the ability to fine tune the
search in the $f$ and $\dot{f}$ directions. We also folded all the
candidates with $\sigma>2$ that appeared in 15 or more observations to
within 0.1\% of their frequency. 0.1\% was chosen because the frequency
modulation caused by the the orbit of 4U~1820$-$30 will not cause $f$ to
deviate by more than 0.1\% (see Section~\ref{sec:tech2}). All the folded
candidates showed noise-like profiles.
%% --------------------------------------------------------
%% -----------------%#############################%--------
%% -----------------%SECTION 3.2 : PHASEMOD SEARCH%--------
%% -----------------%#############################%--------
%% --------------------------------------------------------
\subsection{Phase Modulation Searches}
%% --------------------------------------------------------
\subsubsection{The Search Technique}
\label{sec:tech2}
%% --------------------------------------------------------
In order to analyze longer observations we used the new `phase-modulation'
or `sideband' search technique described by \cite{rce03x}(see also
\citealt{jrs+02}).  This technique relies on the fact that if the
observation time is longer than the orbital period ($T_{\obs}>>P_{\orb}$),
the orbit phase-modulates the pulsar's spin frequency, resulting in a family
of evenly spaced sidebands in the frequency domain, centered around the
intrinsic pulsar signal for circular orbits (see for example
Figure~\ref{figure2}). The width of the region occupied by the sidebands is
$4 \pi f x(T_{\obs}/P_{\orb})$ Fourier bins, where $f$ is the frequency of
the signal and $x$ is the projected semi-major axis. The constant spacing
between the sidebands is $T_{\obs}/P_{\orb}$ Fourier bins.
%% --------------------------------------------------------
\par
%% --------------------------------------------------------
A phase modulation search is conducted by taking short FFTs of portions of
the full power spectrum of an observation. The short FFTs cover overlapping
portions of various lengths (to account for the unknown semi-major axis of
the LMXB) over all the Fourier frequency range of the original power
spectrum (to account for the unknown pulsation period). The secondary power
spectra are then searched for harmonically related peaks indicating
regularly spaced sidebands (to detect the orbital period of the LMXB). Our
phase modulation search included Fourier interpolation and harmonic summing
of up to 4 harmonics of the orbital periods. This particular harmonic
summing accounts for non-circular orbits, but not for non-sinusoidal
pulsations. A detection provides initial estimates of the LMXB spin period,
the orbital period, and the projected radius of the orbit. For interesting
candidate oscillations, these estimates are then refined by generating a
series of complex-valued template responses (i.e. matched filters for the
Fourier domain response of an orbitally modulated signal) to correlate with
the original Fourier amplitudes \citep{rce03x}.
%% --------------------------------------------------------
\subsubsection{Obtaining Candidate Pulsation Frequencies}
\label{sec:cand2}
%% --------------------------------------------------------
To run the phase modulation searches, we first concatenated the orbit-long
lists of events which we had obtained at the end of the preliminary data
preparation into observation-long lists of events. We then binned both the
the observation-long and the orbit-long lists into time series (see
Table~\ref{table2}). We performed an FFT on each, and fed the results into
the local red noise reduction routine. We ran the phase modulation searches
on the original FFTs as well as on the red noise-reduced FFTs. The obtained
lists of candidate oscillations were almost identical: most of the
candidates were millisecond pulsations so the red noise reduction did not
affect them.
%% --------------------------------------------------------
\par
%% --------------------------------------------------------
In order for the phase modulation search to detect a phase modulated signal,
the short FFTs described in Section~\ref{sec:tech2} must overlap at least a
part of the phase modulated signal. When running the phase modulation
searches on the observation-long data sets, we used the range of short FFT
lengths derived in \cite{rce03x} that was appropriate for a wide range of
orbital parameters: from 64 to 131072 bins. As we explained in
Section~\ref{sec:tech2}, the spread of the phase modulated signal in the
Fourier domain is proportional to $xf$ where $x$ is the projected semi-major
axis in light-seconds and $f$ is the intrinsic frequency of the signal.
Because the maximum $x$ for 4U~1820$-$30 is small (for the upper-limit
companion mass of 0.3~\Msun$\;$ and an inclination angle of 90$^{\circ}$,
$x=1.02$~lt-s), the sizes of the short FFTs that we used would allow us to
detect signals up to our Nyquist frequency (2048~Hz). We used a smaller
range of short FFT sizes when we ran the phase-modulation search on the
orbit-long data sets (64 to 32768 bins), and therefore were most sensitive
to signal frequencies $\lesssim$~1~kHz for all possible values of the
companion mass.
%% --------------------------------------------------------
\subsubsection{Candidate Follow-up}
\label{sec:fin2}
%% --------------------------------------------------------
For every candidate oscillation it finds, our phase modulation search
returns, along with other information, a pulsation period, an orbital
period, and $\sigma$. We know from simulation that when a phase modulation
search detects a signal with pulsed fraction equal to or greater than our
upper limit calculated in Section~\ref{sec:sensitivitypm}, the pulsation
period of the signal appears many times in the list of candidates obtained
for that data set due to overlapping numerous short FFTs. Depending on the
strength of the signal, there can also be a jump in the significance
(measured in $\sigma$) between the candidate oscillation corresponding to
the real signal and the candidate with the next smaller significance. When
this happens, if the list of candidates is arranged in order of decreasing
$\sigma$, the real signal appears at the beginning of the list as an outlier
with a singularly large $\sigma$.
%% --------------------------------------------------------
\par
%% --------------------------------------------------------
When examining the phase modulation search candidates, we first extracted
the candidates whose orbital period was 685~s within uncertainties. Among
the extracted candidates, we searched for oscillation periods that appeared
several times in any given observation. There were none. Also among the
extracted candidates, we searched for candidates whose pulsation period
appeared in the candidate lists of more than 10 observations to within a
fractional period of 0.1\%. There were also none. The 0.1\% was a generous
percentage considering that the uncertainties in the candidate pulsation
periods were much smaller. Finally, we searched for outliers within the
candidate lists. We did not find any. 
%% --------------------------------------------------------
%% -------------------%########################%-----------
%% -------------------%SECTION 3.3 : FIT SEARCH%-----------
%% -------------------%########################%-----------
%% --------------------------------------------------------
\subsection{Orbital-Parameter-Fitting Coherent Searches}
%% --------------------------------------------------------
\subsubsection{The Search Technique}
\label{sec:tech3}
%% --------------------------------------------------------
In this search, the source's motion is modelled as a circular orbit.  This
is because LMXB accreting systems are known to have nearly circular
orbits~\citep{zah77}. Hence, we ignored the small errors due to a non-zero
eccentricity. Three parameters are required to specify a particular circular
orbit model: the orbital period $P_{\orb}$, the projected semi--major axis
in light-seconds $x \equiv a \sin (i)/c$ (where $a$ is the real semi-major
axis, $i$ is the inclination of the orbit, and $c$ is the speed of light),
and the orbital phase at the start of an observation $\phi=2 \pi
(T_{\mathrm{start}}-T_{\mathrm{o}})/P_{\orb}$ (where $T_{\mathrm{start}}$ is
the start time of the observation, and $T_{\mathrm{o}}$ is the time of the
last ascending node before $T_{\mathrm{start}}$).  Once a set of trial
orbital parameters is specified, it is straightforward to remove the shifts
that the orbit would introduce in the arrival time of every photon. If the
set of trial orbital parameters matches the real orbital parameters of the
source, the resulting time series will contain the signature of an isolated
pulsar that can be found using a standard pulsation search with no
acceleration.
%% --------------------------------------------------------
\par
%% --------------------------------------------------------
For 4U~1820$-$30, the orbital period is accurately known, but the other two
parameters, $x$ and $\phi$, must be searched. In \cite{ak93}, a maximum
likely value $x_{\max}$ for $x$ is estimated, but the orbital phase is
completely unknown. The sensitivity of detection to errors in the orbital
parameters increases with pulsation frequency (see
Section~\ref{sec:sensitivityFS}), so one must also estimate the highest
likely pulsation frequency $f_{\max}$. From the sensitivity calculations in
Section~\ref{sec:sensitivityFS} we know that the number of sets of trial
orbital parameters increases as $(x_{\max} f_{\max})^2$. For 4U~1820$-$30,
reasonable estimates of these numbers can be provided (see next Section),
and the effort required to search the data is feasible.
%% --------------------------------------------------------
\subsubsection{Obtaining Candidate Pulsation Frequencies}
\label{sec:cand3}
%% --------------------------------------------------------
The method used to run searches by orbital-parameter-fitting is essentially
the same as the standard method for running acceleration searches: the
arrival times of the photons were time-shifted (this time to account for a
{\em{variable}} orbital acceleration), and the resulting data sets were
searched for isolated pulsar signatures. In particular, no matched filtering
in the Fourier domain was done. The current implementation requires about
$2000$~hr to search a typical 4U~1820$-$30 observation on our computers.
This is why it was necessary to limit the number of observations on which to
run the search.
%% --------------------------------------------------------
\par
%% --------------------------------------------------------
\cite{cg01} did a folded period searching of the combined 1996-1997
{\em{RXTE}} standard2 observations of 4U~1820$-$30. They obtained an orbital
period of $P_{orb}$~=~685.0144~$\pm$~0.0054~s. They then did a phase
analysis combining the {\em{RXTE}} 1996-1997 observations of 4U~1820$-$30
with pre-{\em{XTE}} data and found an orbital period of
$P_{\orb}$=685.0119~$\pm$~0.0001~s. The uncertainties in both figures are
small enough (see Section \ref{sec:sensitivityFS}) that we did not need to
search over this parameter.  The decay in $P_{\orb}$ due to emission of
gravitational radiation, estimated to be $\sim$~0.00018~s~yr$^{-1}$
\citep{gravrad}, fell within the above uncertainties. We used
$P_{\orb}~=~685.014$~s when we ran our orbital-parameter-fitting search on
4U~1820$-$30.
%% --------------------------------------------------------
\par
%% --------------------------------------------------------
The maximum plausible $x$, assuming the companion is a degenerate star is
$0.024$~lt-s (this corresponds to a companion mass of 0.078\Msun$\;$ for an
edge-on inclination). With our orbital-parameter-fitting search, we searched
up to an $x_{\max}$ of 0.032~lt-s (corresponding to a companion mass of
0.106\Msun$\;$ for an edge-on inclination. Larger companion masses are
allowed for smaller inclinations). We also set an upper bound on searched
pulsation frequency of $f_{\max}~=~$1250~Hz to save computer time. If a set
of trial orbital parameters corresponds exactly to the real orbital
parameters of 4U~1820$-$30, our orbital-parameter-fitting search collects
successfully all the power due to the signal. If our trial orbital
parameters are slightly off, there will be a sensitivity loss.  When we ran
the search, we tuned the parameter steps to accept up to 50\% loss in signal
power at the maximum frequency. (As frequency decreases or the error in the
size of the orbit decreases, this loss decreases to very near zero; see
Section~\ref{sec:sensitivityFS}).
%% --------------------------------------------------------
\par
%% --------------------------------------------------------
Based on calculations by in \cite{middle}, and using the above known
properties of 4U~1820$-$30 and the above mentioned maximum loss in signal
power, we calculated that the error in a trial $x$ should be no more than
0.000143~lt-s (see Section \ref{sec:sensitivityFS} for details). With that
in mind, we generated an optimal grid of trial orbital parameter sets. The
grid contained several thousands of points (see
Section~\ref{sec:sensitivityFS})
%% --------------------------------------------------------
\par
%% --------------------------------------------------------
For each set of trial orbital parameters, we ran a separate search. We
time-shifted the event data, to reflect the trial orbital parameters.  We
then binned it at a resolution of $0.244$~ms and computed its Fourier
transform.  We applied our local red-noise reduction code and fed the result
to our standard acceleration search code with a maximum acceleration of zero
and an increased significance threshold (i.e.~$\sigma>3$); keeping in mind
that the $\sigma$ returned by the acceleration search after subtracting the
time-shifts is an overestimate of the real $\sigma$ of the signal. This is
because the orbital-parameter-fitting coherent search fails to take into
account the number of times accelerations searches were run. The resulting
list of candidate oscillations was then stored and the process was repeated
for the next set of trial orbital parameters.
%% --------------------------------------------------------
\subsubsection{Candidate Follow-up}
\label{sec:fin3}
%% --------------------------------------------------------
Each candidate returned by the orbital-parameter-fitting code included a
pulsation frequency $f$, an orbital amplitude $x$, an orbital phase $\phi$,
and a significance $\sigma$. Using the returned values for $f$, $x$, and
$\phi$, we folded all the candidates with $\sigma > 5$, allowing the folding
program to search a small distance in the frequency space. Most of the
folded candidates showed noise-like profiles. None of the candidates
corresponding to the more promising-looking folds appeared in more that one
observation except for one candidate. This candidate appeared in two
observations. However, the fold of the second detection of this candidate
showed a noise-like profile.
%% --------------------------------------------------------
\par
%% --------------------------------------------------------
Since folding the statistically significant candidates did not return any
interesting results, we looked for candidates that repeated in more than one
observation (to within a fractional range of
$f_{\mathrm{percent}}$~=~0.001\% and to within a difference in $x$ of
$x_{\mathrm{difference}}=(P_{\orb}*f_{\mathrm{percent}})/(400\pi+2\pi
f_{\mathrm{percent}})$~lt-s). $f_{\mathrm{percent}}$ and
$x_{\mathrm{difference}}$ were derived taking into account the spacing
between the trial values of $x$ and the drop in the amplitude of the
detected signal when the trial orbital parameters were slightly off. There
were about 300 candidates that repeated in 2 observations, and 0 candidates
that repeated in more than 2 observations. We grouped the 300 candidates
into pairs. Assuming the orbital period of $685.0119\pm0.0001$~s reported in
\cite{cg01}, we checked if the orbits of the members of each pair were in
phase with each other. In 18 of the pairs the two orbits were in phase. In
all 18 cases the single trial significances of the members of the pairs were
around 3$\,\sigma$. Again, the folds of all 36 candidates showed noise-like
profiles.
%% --------------------------------------------------------
%% -------------------%#####################%--------------
%% -------------------% SECTION 4 : RESULTS %--------------
%% -------------------%#####################%--------------
%% --------------------------------------------------------
\section{Results and Upper Limits}
\label{sec:results}
%% --------------------------------------------------------
Our searches did not detect pulsations.
%% --------------------------------------------------------
\par
%% --------------------------------------------------------
Since we ran the acceleration searches in a region where they are not
optimized, they were much less sensitive than the other two searches that we
used. Below we explain the procedure we used to set an upper limit on the
pulsed fraction of pulsations from 4U~1820$-$30 detectable with our phase
modulation search technique and our orbital-parameter-fitting coherent
search technique.
%% --------------------------------------------------------
%% ---------------%##################################%-----
%% ---------------%SECTION 4.1 : PHASEMOD SENSITIVITY%-----
%% ---------------%##################################%-----
%% --------------------------------------------------------
\subsection{Phase Modulation Searches}
\label{sec:sensitivitypm}
%% --------------------------------------------------------
The phase modulation searches that we ran on observation-long data sets were
more sensitive than the searches that we ran on the orbit-long data sets
\citep{rce03x}. In this section we present the procedure used to set an
upper limit on the pulsed fraction using the more sensitive search.
%% --------------------------------------------------------
\par
%% --------------------------------------------------------
First, to save computer time, we wrote a modified version of our search
script that searched only a particular region of the FFT. We then ran this
version of the search on several data sets containing simulated pulsations
from 4U~1820$-$30 of various pulsed fractions, for a range of companion
masses (0.02\,\Msun~$\leq M_c \leq$~0.3\,\Msun), and for several spin
frequencies. For the purpose of the simulations, we assumed the known
orbital period of 685~s, a typical {\em{RXTE}} observation duration
(20.5~ks), a count rate of 2050 counts per second, an inclination angle of
60$^{\circ}$, and a circular orbit. Figure~\ref{figure3} shows the results
of the simulations for this type of search run on data sets containing fake
3~ms sinusoidal pulsations. Every point in the figure corresponds to 100
searched data sets. The shading of every point corresponds to the detected
power of the signal. The power was then converted to probability and
multiplied by the number of independent trials (roughly the original number
of time bins) to convert it to $\sigma$. In Figure~\ref{figure3}, the line
of $\sigma$ = 5 indicates the value of our 95\% confidence upper limit on
the pulsed fraction for a signal period of 3~ms. This value varied between
0.55\% and 1.0\% (background unsubtracted) over the range of companion
masses that we used.  The value was unaffected by the background correction
since the background was small compared to the source count rate: using the
`FTOOL' \texttt{PCABACKEST}, we estimated the background count to be 10
counts per second per PCU. This is only 2\% of the typical count rate of the
source. This means that if there were a 3-ms coherent pulsation coming from
4U~1820$-$30 with a pulsed fraction equal to or higher than the upper limit
stated above, our searches would have detected it at $\sigma \geq 5\,$ 95\% of
the time. The upper limit on the pulsed fraction decreased slightly when the
period of the signal was larger by a few milliseconds. This is because, at
longer spin periods, the width of the sidebands feature is smaller (see
Section~\ref{sec:tech2}); and hence the power accumulated in each of the
sidebands is greater. For example, for a 10-ms coherent pulsation, this
value varied between 0.3\% and 0.5\% over the range of companion masses that
we used. Similarly, the upper limit on the pulsed fraction increases for
smaller spin frequencies.
%% --------------------------------------------------------
\par 
%% --------------------------------------------------------
Our upper limit on the pulsed fraction of possible signals from 4U~1820$-$30
is $\sim$0.8\% for spin periods close to 3~ms. 
The pulsed fractions of the signals detected from the 5 LMXBs with known
signal period ranged between 3\% and 7\%, for signal frequencies between 2
and 6~ms \citep{source1,source2, source3, source4,
source4sum}.
%% --------------------------------------------------------
%% ---------------%###################################%----
%% ---------------%SECTION 4.2 : FITSEARCH SENSITIVITY%----
%% ---------------%###################################%----
%% --------------------------------------------------------
\subsection{Orbital-Parameter-Fitting Coherent Searches}
\label{sec:sensitivityFS}
%% --------------------------------------------------------
The goal of orbital-parameter-fitting is to achieve a sensitivity close to
the sensitivity one would have if the pulsar were isolated.  To achieve
this, one must compute how densely to sample the three-dimensional space of
orbital parameters.  Such calculations have already been done by
\cite{middle}, but in this section we work through them with more details to
better understand the effects of small errors in the trial orbital period.
%% --------------------------------------------------------
\par
%% --------------------------------------------------------
First, assume that $P_{\orb}$ is known exactly.  Then combine $x$ and $\phi$
into a single complex number $X$
%% --------------------------------------------------------
\[
X=xe^{i\phi}.
\]
%% --------------------------------------------------------
If the pulsar emits sinusoidal pulsations at a frequency $f$, then the
received signal will be
%% --------------------------------------------------------
\[
s(t) = \sin(2\pi f\left(t + Re\left(X_{\orb} e^{2\pi i t/P_{\orb}}\right)
\right)),
\]
%% --------------------------------------------------------
up to a phase shift (which we will ignore).  If the pulse profile is not
sinusoidal, this transformation will happen equally to all harmonics; we
will ignore this too.
%% --------------------------------------------------------
\par
%% --------------------------------------------------------
Suppose that we remove the effects of the orbital motion by time-shifting
the incoming events according to a model using the known $P_{\orb}$ but $X =
X_{\mathrm{trial}}$.  The observed signal will then be
%% --------------------------------------------------------
\begin{eqnarray*}
s(t)
& = & \sin(2\pi f(t + \\
&   & Re\left( X_{\orb} e^{2\pi i t/P_{\orb}}\right) - 
      Re\left( X_{\trial} e^{2\pi i t/P_{\orb}}\right)  )) \\
& = & \sin(2\pi f(t + \\
&   & Re\left( (X_{\orb}-X_{\trial}) e^{2\pi i t/P_{\orb}}\right))).
\end{eqnarray*}
%% --------------------------------------------------------
In other words, the effect will be exactly that of a residual phase
modulation of complex amplitude $X_{\orb}-X_{\trial}$.
%% --------------------------------------------------------
\par
%% --------------------------------------------------------
Such a phase modulation produces a comb-like pattern in the Fourier domain
(see Figure~\ref{figure2}). For narrow combs, the tallest peak is the
central peak: the one at the true pulsation frequency. The amplitude of the
central peak falls off as $J_0(\left| X_{\orb}-X_{\trial}\right| 2 \pi f)$
where $J_0$ is a 0$^{th}$ order Bessel function \citep{rce03x,middle}.
Therefore, the sensitivity loss (in power) due to a residual phase
modulation of size $\left|X_{\orb}-X_{\trial}\right| = \varepsilon$ will be
$J_0(\varepsilon 2 \pi f)^2$. This means that if we search the $X_{\orb}$
space with a sufficient density to fall within $\varepsilon$ of every
plausible $X_{\orb}$, we will detect the source with a sensitivity only a
factor of $J_0(\varepsilon 2 \pi f)^2$ less than that we would have for an
isolated pulsar. In other words, the residual time shifts need to be small
compared to the pulsation period.
%% --------------------------------------------------------
\par
%% --------------------------------------------------------
%%%%%%%% SOME SENSITIVITY NOTES:
%%%%%%%% important number 0.000143412~ms. 
%%%%%%%% This number (0.000143) was obtained from the sens cancluations at fmax
%%%%%%%% the size of the region searched in the x splace is a circle of radius xmax
%%%%%%%% therefore the number of orbital paramater trials increases as $x_{\max}
%%%%%%%% f_{\max}$.
%%%%%%%% spacing between the points of the grid is (2*cos(pi/8)*maxres = 0.000264).
%%%%%%%% (we actually used a spacing of 0.000277 and that leads to xresmax 0.0001499
%%%%%%%% which lead to a max loss of 0.4656).
%% --------------------------------------------------------
For this particular situation, we fixed the maximum $\left|X\right|$ as
0.032~lt-ms, $f_{\max}$ as 1250~Hz, and the sensitivity factor as 0.5 at
$f_{\max}$. The maximum allowable residual $X$ (which we called
$\varepsilon$) is then 0.000143~ms. With that in mind, we generated an
optimal grid of trials for complex-valued $X$. It turns out that the optimal
grid (the one where the parameter space is searched most efficiently) is a a
circle-packing array: a grid where the values of $X_{\trial}$ are
distributed on the corners of equilateral triangles \citep[e.g.,][]{middle}.
The point in the middle of every triangle is the point that is furthest away
from the closest $X_{\trial}$ and therefore the distance between it and the
closest $X_{\trial}$ should be $\varepsilon$~=~0.000143~lt-s. The spacing
between any two guesses is then 0.000264~lt-s. The total number of trials of
$X$ is then the area of the complex $X$-plane containing the trials (roughly
$\pi {x_{\max}}^2$) divided by the area covered by every $X_{\trial}$
(roughly $\pi {\varepsilon}^2$). This amounts to an approximate number of
50000 trials per observation. With the trials of $X$ distributed as we
described, our worst case sensitivity to a given signal of frequency $f$ is
then ${J_0(0.000143 \times 2 \pi f)^2}$ $\times$ (the sensitivity to an
isolated pulsar signal of frequency $f$). At $f=f_{\max}$ the value of this
factor is 0.5. Its value increases for signals of lower frequency. For
$f={f_{\max}}/2$, it is 0.85.
%% --------------------------------------------------------
\par
%% --------------------------------------------------------
In the above calculation, we assumed no loss of power due to errors in the
assumed orbital period. But how accurately {\emph{does}} $P_{\orb}$ need to
be known? We can do an approximate calculation:
suppose that $P_{\trial} = (1-\epsilon)P_{\orb}$. The resulting signal
is then
%% --------------------------------------------------------
\begin{eqnarray*}
s(t)
& = & \sin(2\pi f(t + \\
&   & Re\left( X_{\orb} e^{2\pi i t/P_{\orb}}\right) - 
      Re\left( X_{\trial} e^{2\pi i t/P_{\trial}}\right) ))\\
& \simeq & \sin(2\pi f(t + \\
&   & Re\left( (X_{\orb}-X_{\trial} e^{2\pi i t \epsilon/P_{\orb}} ) 
      e^{2\pi i t/P_{\orb}}\right) )).
\end{eqnarray*}
%% --------------------------------------------------------
This amounts to an additional slowly-varying error in the estimation of the
orbital phase $\phi$.  Errors in $\phi$ that are independent of
($P_{\orb}-P_{\trial}$) have already been taken into account by the earlier
sensitivity calculations. Therefore, when we calculated the effects of the
additional slowly-varying error in $\phi$ in a given observation of length
$T_{\obs}$, we could assume, to start with, that $\phi$ is correct in
the middle of that observation. At the end of that observation, we would
have accumulated an error of size $\epsilon \pi T_{\obs}/P_{\orb}$ in the
estimation of $\phi$. This adds an {\emph{additional}} residual $X$ of about
$\epsilon \left|X\right| \pi T_{\obs}/P_{\orb}$.  In the worst case, this is
about $4.3 \epsilon$~lt-s. From the difference between the value we used for
the orbital period (685.014~s) and the lower bound on the more accurate
orbital period measured in \cite{cg01} of 685.0119$\pm$0.0001~s, we estimate
$\epsilon \sim$~0.000003. Hence the size of the maximum additional residual
$X$ is $\sim$~0.000013~lt-s, which is a factor of $\sim$11 smaller than the
maximum residual $X$ that the spacing in our grid of trials is allowing. In
the worst case scenario (at $f=f_{\max}$ and $x=x_{\max}$), if we combine
the maximum error in $X$ due to the spacing of our trials with the
additional error in $X$ due to the error in the orbital period, we find a
worst case sensitivity of 0.43~$\times$~that for an isolated pulsar (instead
of 0.5 when the $P_{\orb}$ is exactly right). If the error caused by the
uncertainty in $P_{\orb}$ were slightly larger, it would have caused an even
larger decrease in the detectable power, and we would have been forced to
search in $P_{\orb}$ also.
%% --------------------------------------------------------
\par
%% --------------------------------------------------------
Above, we have extracted formulae for the sensitivity of the
orbital-parameter-fitting search in terms of the sensitivity of searches for
an isolated pulsar. These, as well as a small number of simulations, show
the following: if there are no errors in the search parameters, we can
detect at 3$\,\sigma$ a signal of pulsed fraction 0.17\%, and at
5.3$\,\sigma$ a signal of pulsed fraction 0.2\%. If our trials for the
orbital parameters (including the orbital period) are off by the worst-case
amounts, we can detect at 3$\,\sigma$ a signal of pulsed fraction 0.26\%,
and at 5.3$\,\sigma$ a signal of pulsed fraction 0.3\%.  The above numbers
are summarized in Table~\ref{table3}. Because of the way we are doing the
candidate follow-up (i.e.~folding candidates that appear several times in
the list of candidates), the occurrence of the signal in more than one
observation increases greatly the chance of detecting it: if a signal
appeared only in one observation at 3$\,\sigma$, we would have discarded it
as a false candidate. If a signal appeared in several observations at
3$\,\sigma$ we would certainly not have missed it. Finally, any individual
signal that appeared at $\sigma \geq 5.3$ would have been detected with
confidence.
%% --------------------------------------------------------
\par
%% --------------------------------------------------------
To summarize, our sensitivity formulae and our simulations show reliable
detection at a pulse fraction of $0.3$\% for the parameters we have chosen.
This is of course dependent on a number of assumptions: the companion must
be a degenerate star, the pulsation frequency must be less than about
1250~Hz, and $P_{\orb}$ must be within 0.0022~ms of 685.014~ms. It should be
noted, however, that these assumptions are covariant: if the pulsation
frequency is lower, the sensitivity improves.  If $x$ is smaller than the
assumed upper bound, we become less sensitive to an error in $P_{\trial}$.
%% --------------------------------------------------------
%% --------------------%######################%------------
%% --------------------%SECTION 5 : DISCUSSION%------------
%% --------------------%######################%------------
%% --------------------------------------------------------
\section{Discussion}
\label{sec:discussion}
%% --------------------------------------------------------
The search described in this paper represents the first source in a larger
program to search for coherent pulsations in the persistent LMXBs.  There
are currently five known accreting MSPs, all of which have pulse frequencies
in the range of 185 to 435~Hz and low luminosities.  The pulsed fractions
observed are in the 3~--~7~\% range. Recently, observations of burst
oscillations, kHz QPOs, and coherent pulsations in two of these accreting
MSPs have confirmed that the burst oscillations are indeed at the spin
frequency (\citealt{THEnaturepaper}; \citealp*{secondproof}) and that the
kHz~QPO separation can be either near the spin frequency or half the spin
frequency. Thus, there is now a total sample of 14 LMXBs with
well-determined spin frequencies. In 5 LMXBs, coherent pulsations are
observed \citep{source1,source2,source3,source4,source5}. In 11 LMXBs burst
oscillations are observed \citep{str01d,1814paper}. In 2 of these LMXBs,
both coherent pulsations and burst oscillations are observed
\citep{THEnaturepaper,1814paper}.
%% --------------------------------------------------------
\par
%% --------------------------------------------------------
We can see pulsations during X-ray bursts and in the non-burst emission in a
handful of transient sources. Why then can we not see pulsations in the
persistent emission of the many bright LMXBs including 4U~1820$-$30?
%% --------------------------------------------------------
\par
%% --------------------------------------------------------
For any individual pulsar, an unfavorable viewing geometry or rotational
geometry can prevent the pulsations from being seen. However, a systematic
absence of pulsations from all persistent LMXBs would require a stronger
explanation. There are three main categories of explanations that have been
considered. These are: relativistic effects that decrease the observed
pulsed fraction, smearing of pulses from the surface due to electron
scattering, and surface magnetic fields that are too weak to significantly
channel the accretion flow (possibly as a result of magnetic screening). We
briefly consider each of the possibilities below.
%% --------------------------------------------------------
\par
%% --------------------------------------------------------
%% LENSING ------------------------------------------------
%% --------------------------------------------------------
In \citet*{lensing1}, the authors demonstrate that gravitational
light-bending effects on emission from hot polar caps can greatly reduce the
pulsed fraction: the pulses are distorted and the pulsed fraction decreases
as a star becomes more relativistic because its radius is decreasing.  The
decrease in the pulsed fraction results from the fact that the gravitational
deflection of photons displaces the horizon away from the observer. In other
words, more of the star than a hemisphere is visible to the observer. Thus,
one of the caps is always in nearly full view as the other is disappearing.
It is possible for the pulsed fraction of a given source to be reduced down
to $\sim$1\%, provided the radius of the emitting star is small enough and
the opening angle of the polar caps is large enough \citep*{lensing2}.  This
fraction is larger than our upper limit on the pulsed fraction, suggesting
that gravitational lensing effects may not be the only factor preventing us
from seeing the pulsations. Also, the five known accreting MSPs exhibit
pulsed fractions of more than 5\%. There is no reason to believe that those
systems have a significantly different value of $M/R$ than the persistent
LMXBs. Consequently, it seems that the relativistic effects are not the
primary explanation for the small pulsed fraction in the latter sources.

%% --------------------------------------------------------
\par
%% --------------------------------------------------------
%% SCATTERING ---------------------------------------------
%% --------------------------------------------------------
Another explanation for low modulation fractions, electron scattering, has
been revisited by \citet*{THEwhypaper}.  The existence of a corona (a
spherical cloud) of hot electrons surrounding the accreting star has been
inferred from the analysis of the X-ray spectra of several LMXBs
\citep[e.g.,][]{scatpaper,THEwhypaper}.  This hot corona is
presumably upscattering softer seed photons and smearing out the pulsations.
\cite{THEwhypaper} argue that it is because of the large optical depth of
this corona that pulsations have not been detected from most LMXBs.  They
claim that the optical depth below which millisecond pulsations become
detectable is $\tau\sim4$. They also do spectral fits for several LMXBs and
extract the parameter $\tau$. For the majority of the analysed LMXBs they
find that the optical depth of the Compton cloud is at least twice as high
as $\tau=4$. For these LMXBs, they claim that any high frequency pulsations
(higher than $\sim$300~Hz) would be below detectability.
\citeauthor{THEwhypaper} also make the observation that there is a
possibility of finding pulsations in the sources that are much less
luminous, because in these sources, the spherical Compton cloud is more
transparent ($\tau<4$). Unfortunately, the smallest reported $\tau$ for the
spherical Compton cloud that is presumably surrounding 4U~1820$-$30 was
$\tau=6$ \citep{spec1}.  In all the other papers that discuss spectra from
4U~1820$-$30, the authors used a Comptonization model with a black body
component to fit the spectra and found that $\tau$ falls between 11 and 13
\citep*[e.g.,][]{spec2}. If what \cite{THEwhypaper} have claimed is
true, the large optical depth of the Comptonizing cloud could very well be
the reason we are not seeing pulsations from 4U~1820$-$30.
%% --------------------------------------------------------
\par
%% --------------------------------------------------------
%% SCREENING ----------------------------------------------
%% --------------------------------------------------------
The possibility of magnetic screening by accreted material at high accretion
rates has been examined by \citet*{magpaper}.  Their reasoning goes as
follows: to get preferential heating of the stellar surface, the magnetic
field must be large enough to disrupt the accretion flow and channel it onto
the polar caps. Preferential heating of the surface causes the observed
emission to be pulsed. If the accretion flow is not disrupted by a
magnetosphere, the accretion disk is believed to extend almost to the
stellar surface and pulsations are no longer seen. This could happen because
the external dipole magnetic field of the star is `buried' or `screened' by
the accreted matter. \cite{magpaper} study the `burial' scenario.  In
particular, they study the competition between downward compression of the
magnetic flux by accretion at a given $\dot{M}$, and its upward transport by
diffusion.  They find that the diffusion and accretion time scales are equal
for $\dot{M}\sim0.1{\dot{M}}_{\mathrm{Edd}}$, where
${\dot{M}}_{\mathrm{Edd}}$ is the accretion rate given by the Eddington
limit. This means that for larger $\dot{M}$ there is little chance that
magnetic flux makes it to the surface and therefore little chance of
observing pulsations. They also find that magnetic screening is completely
ineffective for $\dot{M}<0.02{\dot{M}}_{\mathrm{Edd}}$, so that, no matter
how the accreted material joins onto the star, the underlying stellar field
would always be evident.  The typical range of mass accretion rates in LMXBs
is $\dot{M}\geq0.01{\dot{M}}_{\mathrm{Edd}}$
($\dot{M}\geq10^{-10}~\Msun~$yr$^{-1}$) \citep{THEnaturepaper}.
\cite{magpaper} propose that the absence of persistent millisecond
pulsations from most neutron stars in LMXBs is due to their high accretion
rates. Discoveries made since then are consistent with their hypothesis: all
five persistent millisecond X-ray pulsars are low-luminosity transients in
very close binaries that lie at the low end of the mean $\dot{M}$
distribution for LMXBs with $\dot{M}\sim10^{-11}~\Msun~$yr$^{-1}$
\citep{source3, source4sum, THEnaturepaper}. It is tempting to adopt
magnetic screening as the main explanation for the lack of observed
pulsations from 4U~1820$-$30 because the luminosity of this source
($\sim4\times{10}^{37}$~ergs/s) is on the high end of luminosities of atoll
sources, implying a normal (rather than a low) accretion rate of the
in-falling matter. \cite{spw87} estimated this accretion rate to be
$\sim5\times10^{-9}$~\Msun~yr$^{-1}$.
%% --------------------------------------------------------
\par
%% --------------------------------------------------------
%% --------------------%###################%---------------
%% --------------------%SECTION 6 : SUMMARY%---------------
%% --------------------%###################%---------------
%% --------------------------------------------------------
\section{Summary}
\label{sec:sum}
\begin{figure}
%% emulateapj command  >>>>>> \includegraphics[scale=.5]{figure1.eps}
%# manuscript command  >>>>>> \includegraphics[scale=.8]{figure1.eps}
\includegraphics[scale=.8]{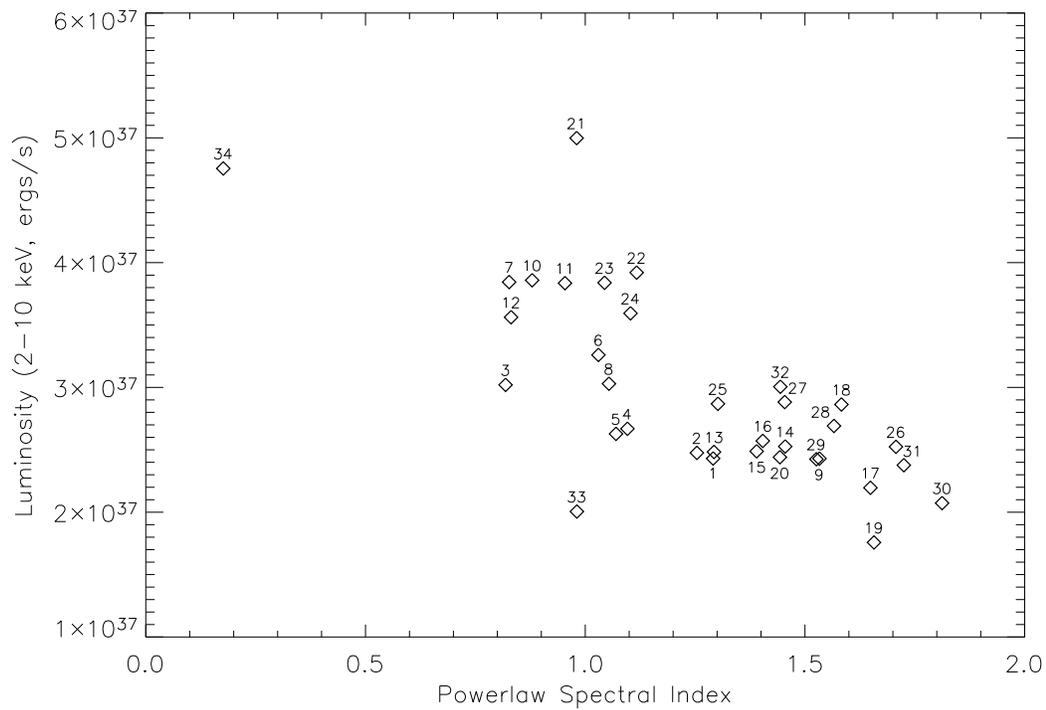}
%% --------------------------------------------------------
\caption
{
The power-law spectral index and the 2--10~keV luminosity of each of the 34
observations that we have searched for pulsations. These numbers were
obtained from spectral fittings and were based on an assumed distance of
6.4~kpc.
\label{figure1}
}
\end{figure}

\begin{figure}
%% emulateapj command  >>>>> \includegraphics[angle=270,scale=.34]{figure2.eps}
%% manuscript command  >>>>> \includegraphics[angle=270,scale=.6]{figure2.eps}
\includegraphics[angle=270,scale=.6]{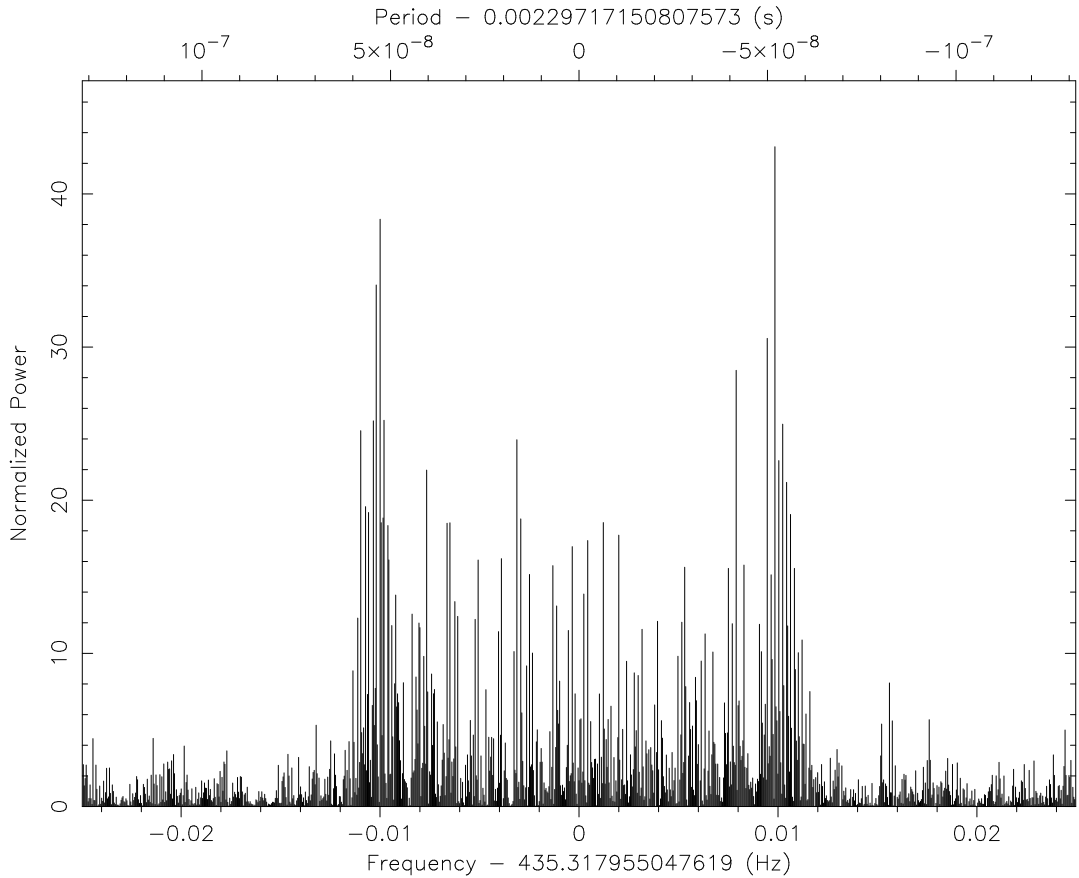}
%% --------------------------------------------------------
\caption
{
Sidebands in the FFT of observation 70131-01-05-00 (4 {\em{RXTE}} orbits
long) of XTE~J1751-305 centered on the intrinsic pulsar spin
frequency (435.318~Hz) and caused by orbital phase modulations of the pulsar
frequency (see Section~\ref{sec:tech2}). The binary orbital period that is
causing the above modulation is $\sim$42.4~min and the projected
orbital radius is $\sim$10.11~lt-ms. The pulsed fraction above is high
(3.76\%, background unsubtracted) which makes the signal stand well above
the noise. The width of the sideband pattern is related to the projected
semi-major axis of the orbit. The above signal was detected at
10.6$\,\sigma$ using our phase modulation search technique and at
10.8$\,\sigma$ using our acceleration search technique.
\label{figure2}
}
\end{figure}

\begin{figure}
%% emulateapj command >>>>> \includegraphics[scale=.45]{figure3.eps}
%% manuscript command >>>>> \includegraphics[scale=.8]{figure3.eps}
\includegraphics[scale=.8]{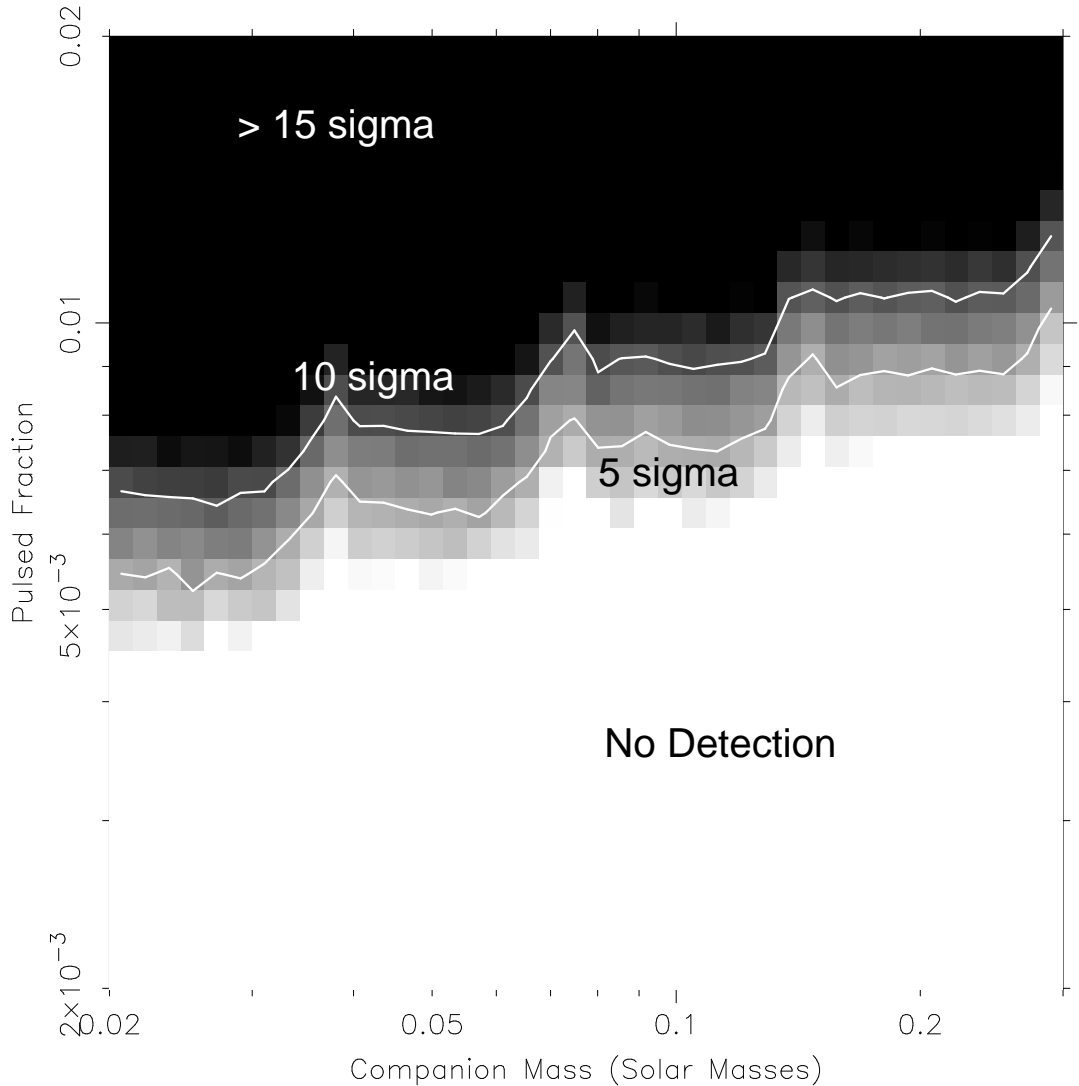}
%% --------------------------------------------------------
\caption
{
Monte-Carlo derived sensitivity calculations for the pulsation
searches of archival {\em{RXTE}} observations of 4U~1820$-$30. The plotted
sensitivities are 95\% confidence limits using the phase modulation search
technique for a typical {\em{RXTE}} observation assuming a pulsar spin
period of
3~ms. With these types of searches,
we would easily have detected any coherent pulsations with a pulsed
fraction~$\gtrsim$~0.8\% for all realistic companion masses and spin periods.
\label{figure3}
}
\end{figure}

%% --------------------------------------------------------
In this paper, we have described a search of 4U1820$-$30 that set upper
limits on the pulsed fraction covering essentially all plausible pulse
periods and companion masses for this system.  These limits were between
0.3\% and 0.8\% over this whole range. We also searched a broad range of
luminosity and spectral states (see Figure~\ref{figure1}).  The luminosities
covered the range from 0.08 to 0.25 of the Eddington luminosity for a
1.4~\Msun$\;$ neutron star and a distance of 6.4~kpc. 4U1820$-$30 is
interesting in this regard because the luminosity varies by a factor of
three. 
%% --------------------------------------------------------
\par
%% --------------------------------------------------------
For 4U~1820$-$30, our upper limits are about an order of magnitude lower in
pulsed fraction than the known accreting MSPs. However, this is only one
source and any number of specific features, such as unfavorable viewing or
rotational geometries, in addition to electron scattering and magnetic
screening, could prevent the pulsations from being seen. The most
interesting conclusions will be drawn once we have a sample of many more
sources that have been searched to similarly deep levels.
%% --------------------------------------------------------
\par
%% --------------------------------------------------------
%% ---------------%####################%-------------------
%% ---------------%END OF THE MAIN TEXT%-------------------
%% ---------------%####################%-------------------
%% --------------------------------------------------------
%% -------------------%###############%--------------------
%% -------------------%ACKNOWLEDGMENTS%--------------------
%% -------------------%###############%--------------------
%% --------------------------------------------------------
\acknowledgments
%% --------------------------------------------------------
This work was supported by the Natural Sciences and Engineering Research
Council (NSERC) Julie Payette research scholarship. It would also have been
impossible to do this work without the computing power of the Beowulf
computer cluster operated by the McGill University Pulsar Group and funded
by the Canada Foundation for Innovation. Additional support came from NSERC,
FQRST, CIAR, and NASA. Victoria M. Kaspi is a CRC Chair and NSERC Steacie
Fellow.
%% --------------------------------------------------------
%% -------------------%############%-----------------------
%% -------------------%BIBLIOGRAPHY%-----------------------
%% -------------------%############%-----------------------
%% --------------------------------------------------------
%% \bibliographystyle{apj}
%% \bibliography{journapj,psrrefs,pulsars,rimpsr}

\begin{thebibliography}{45}
\expandafter\ifx\csname natexlab\endcsname\relax\def\natexlab#1{#1}\fi

\bibitem[{{Anderson} {et~al.}(1997){Anderson}, {Margon}, {Deutsch}, {Downes},
  \& {Allen}}]{amd+97}
{Anderson}, S.~F., {Margon}, B., {Deutsch}, E.~W., {Downes}, R.~A., \& {Allen},
  R.~G. 1997, \apjl, 482, L69+
  
\bibitem[{{Arons} \& {King}(1993)}]{ak93}
{Arons}, J. \& {King}, I.~R. 1993, \apjl, 413, L121
  
\bibitem[{{Bloser} {et~al.}(2000){Bloser}, {Grindlay}, {Kaaret}, {Zhang},
  {Smale}, \& {Barret}}]{spec1}
{Bloser}, P.~F., {Grindlay}, J.~E., {Kaaret}, P., {Zhang}, W., {Smale}, A.~P.,
  \& {Barret}, D. 2000, \apj, 542, 1000

\bibitem[{{Campana} {et~al.}(2003){Campana}, {Ravasio}, {Israel}, {Mangano}, \&
  {Belloni}}]{source4sum}
{Campana}, S., {Ravasio}, M., {Israel}, G.~L., {Mangano}, V., \& {Belloni}, T.
  2003, \apjl, in preparation.
  
\bibitem[{{Chakrabarty} {et~al.}(2003){Chakrabarty}, {Morgan}, {Muno},
  {Galloway}, {Wijnands}, {van der Klis}, \& {Markwardt}}]{THEnaturepaper}
{Chakrabarty}, D., {Morgan}, E.~H., {Muno}, M.~P., {Galloway}, D.~K.,
  {Wijnands}, R., {van der Klis}, M., \& {Markwardt}, C.~B. 2003, \nat, 424, 42
  
\bibitem[{{Chou} \& {Grindlay}(2001)}]{cg01}
{Chou}, Y. \& {Grindlay}, J.~E. 2001, \apj, 563, 934

\bibitem[{{Cumming} {et~al.}(2001){Cumming}, {Zweibel}, \&
  {Bildsten}}]{magpaper}
{Cumming}, A., {Zweibel}, E., \& {Bildsten}, L. 2001, \apj, 557, 958

\bibitem[{{Dotani} {et~al.}(1989){Dotani}, {Mitsuda}, {Makishima}, \&
  {Jones}}]{dmm+89}
{Dotani}, T., {Mitsuda}, K., {Makishima}, K., \& {Jones}, M.~H. 1989, \pasj,
  41, 577
  
\bibitem[{{Galloway} {et~al.}(2002){Galloway}, {Chakrabarty}, {Morgan}, \&
  {Remillard}}]{source3}
{Galloway}, D.~K., {Chakrabarty}, D., {Morgan}, E.~H., \& {Remillard}, R.~A.
  2002, \apjl, 576, L137

\bibitem[{{Groth}(1975)}]{gro75}
{Groth}, E.~J. 1975, \apjs, 29, 285

\bibitem[{Johnston \& Kulkarni(1991)}]{jk91}
Johnston, H.~M. \& Kulkarni, S.~R. 1991, ApJ, 368, 504

\bibitem[{{Jouteux} {et~al.}(2002){Jouteux}, {Ramachandran}, {Stappers},
  {Jonker}, \& {van der Klis}}]{jrs+02}
  {Jouteux}, S., {Ramachandran}, R., {Stappers}, B.~W., {Jonker}, P.~G., \& {van
  der Klis}, M. 2002, \aap, 384, 532

\bibitem[{{Kaaret} {et~al.}(1999){Kaaret}, {Piraino}, {Bloser}, {Ford},
  {Grindlay}, {Santangelo}, {Smale}, \& {Zhang}}]{kpb+99}
{Kaaret}, P., {Piraino}, S., {Bloser}, P.~F., {Ford}, E.~C., {Grindlay}, J.~E.,
  {Santangelo}, A., {Smale}, A.~P., \& {Zhang}, W. 1999, \apjl, 520, L37

\bibitem[{{Kylafis} \& {Klimis}(1987)}]{scatpaper}
{Kylafis}, N.~D. \& {Klimis}, G.~S. 1987, \apj, 323, 678

\bibitem[{{Markwardt} {et~al.}(2003{\natexlab{a}}){Markwardt}, {Smith}, \&
  {Swank}}]{source4}
{Markwardt}, C.~B., {Smith}, E., \& {Swank}, J.~H. 2003{\natexlab{a}},
  \iaucirc, 8080, 2

\bibitem[{{Markwardt} {et~al.}(2003{\natexlab{b}}){Markwardt}, {Strohmayer}, \&
  {Swank}}]{secondproof}
{Markwardt}, C.~B., {Strohmayer}, T.~E., \& {Swank}, J.~H. 2003{\natexlab{b}},
  The Astronomer's Telegram, 164, 1

\bibitem[{{Markwardt} \& {Swank}(2003)}]{source5}
{Markwardt}, C.~B. \& {Swank}, J.~H. 2003, \iaucirc, 8144, 1

\bibitem[{{Markwardt} {et~al.}(2002){Markwardt}, {Swank}, {Strohmayer}, {Zand},
  \& {Marshall}}]{source2}
{Markwardt}, C.~B., {Swank}, J.~H., {Strohmayer}, T.~E., {Zand}, J.~J.~M.~i.,
  \& {Marshall}, F.~E. 2002, \apjl, 575, L21

\bibitem[{{Meszaros} {et~al.}(1988){Meszaros}, {Riffert}, \&
  {Berthiaume}}]{lensing2}
{Meszaros}, P., {Riffert}, H., \& {Berthiaume}, G. 1988, \apj, 325, 204

\bibitem[{{Middleditch}(2004)}]{middle}
{Middleditch}, J. 2004, ArXiv Astrophysics e-prints

\bibitem[{{Priedhorsky} \& {Terrell}(1984)}]{pt84}
{Priedhorsky}, W. \& {Terrell}, J. 1984, \apjl, 284, L17

\bibitem[{{Ransom} {et~al.}(2003){Ransom}, {Cordes}, \& {Eikenberry}}]{rce03x}
{Ransom}, S.~M., {Cordes}, J.~M., \& {Eikenberry}, S.~S. 2003, \apj, 589, 911

\bibitem[{{Ransom} {et~al.}(2002){Ransom}, {Eikenberry}, \&
  {Middleditch}}]{rem02x}
{Ransom}, S.~M., {Eikenberry}, S.~S., \& {Middleditch}, J. 2002, \aj, 124, 1788

\bibitem[{{Rappaport} {et~al.}(1987){Rappaport}, {Ma}, {Joss}, \&
  {Nelson}}]{rmj+87}
{Rappaport}, S., {Ma}, C.~P., {Joss}, P.~C., \& {Nelson}, L.~A. 1987, \apj,
  322, 842
  
\bibitem[{{Sansom} {et~al.}(1989){Sansom}, {Watson}, {Makishima}, \&
  {Dotani}}]{sw89}
{Sansom}, A.~E., {Watson}, M.~G., {Makishima}, K., \& {Dotani}, T. 1989, \pasj,
  41, 591
  
\bibitem[{{Smale} {et~al.}(1987){Smale}, {Mason}, \& {Mukai}}]{smm86}
{Smale}, A.~P., {Mason}, K.~O., \& {Mukai}, K. 1987, \mnras, 225, 7P

\bibitem[{{Smale} {et~al.}(1997){Smale}, {Zhang}, \& {White}}]{szw97}
{Smale}, A.~P., {Zhang}, W., \& {White}, N.~E. 1997, \apjl, 483, L119+

\bibitem[{Stella {et~al.}(1987)Stella, Priedhorsky, \& White}]{spw87}
Stella, L., Priedhorsky, W., \& White, N.~E. 1987, ApJ, 312, L17

\bibitem[{{Stella} {et~al.}(1987){Stella}, {White}, \& {Priedhorsky}}]{swp87}
{Stella}, L., {White}, N.~E., \& {Priedhorsky}, W. 1987, \apjl, 315, L49

\bibitem[{{Strohmayer}(2001)}]{str01d}
{Strohmayer}, T.~E. 2001, Advances in Space Research, 28, 511

\bibitem[{{Strohmayer} \& {Brown}(2002)}]{sb02}
{Strohmayer}, T.~E. \& {Brown}, E.~F. 2002, \apj, 566, 1045

\bibitem[{{Strohmayer} {et~al.}(2003){Strohmayer}, {Markwardt}, {Swank}, \&
  {in't Zand}}]{1814paper}
{Strohmayer}, T.~E., {Markwardt}, C.~B., {Swank}, J.~H., \& {in't Zand}, J.
  2003, \apjl, 596, L67

\bibitem[{{Tan} {et~al.}(1991){Tan}, {Morgan}, {Lewin}, {Penninx}, {van der
  Klis}, {van Paradijs}, {Makishima}, {Inoue}, {Dotani}, \& {Mitsuda}}]{tlp91}
{Tan}, J., {Morgan}, E., {Lewin}, W.~H.~G., {Penninx}, W., {van der Klis}, M.,
  {van Paradijs}, J., {Makishima}, K., {Inoue}, H., {Dotani}, T., \& {Mitsuda},
  K. 1991, \apj, 374, 291

\bibitem[{{Titarchuk} {et~al.}(2002){Titarchuk}, {Cui}, \&
  {Wood}}]{THEwhypaper}
{Titarchuk}, L., {Cui}, W., \& {Wood}, K. 2002, \apjl, 576, L49

\bibitem[{{van der Klis}(1999)}]{QPOreview}
{van der Klis}, M. 1999, in Pulsar Timing, General Relativity and the Internal
  Structure of Neutron Stars, 259--+
  
\bibitem[{van~der Klis {et~al.}(1993)van~der Klis, Hasinger, Dotani, Mitsuda,
  Verbunt, Murphy, van Paradijs, Belloni, Makishima, Morgan, \& Lewin}]{vhd+93}
van~der Klis, M., Hasinger, G., Dotani, T., Mitsuda, K., Verbunt, F., Murphy,
  B.~W., van Paradijs, J., Belloni, T., Makishima, K., Morgan, E., \& Lewin, W.
  H.~G. 1993, MNRAS, 260, 686
  
\bibitem[{{van der Klis} {et~al.}(1993){van der Klis}, {Hasinger}, {Verbunt},
  {van Paradijs}, {Belloni}, \& {Lewin}}]{vhv+93}
{van der Klis}, M., {Hasinger}, G., {Verbunt}, F., {van Paradijs}, J.,
  {Belloni}, T., \& {Lewin}, W.~H.~G. 1993, \aap, 279, L21
  
\bibitem[{{Vaughan} {et~al.}(1994){Vaughan}, {van der Klis}, {Wood}, {Norris},
  {Hertz}, {Michelson}, {van Paradijs}, {Lewin}, {Mitsuda}, \&
  {Penninx}}]{vvw+94x}
{Vaughan}, B.~A., {van der Klis}, M., {Wood}, K.~S., {Norris}, J.~P., {Hertz},
  P., {Michelson}, P.~F., {van Paradijs}, J., {Lewin}, W. H.~G., {Mitsuda}, K.,
  \& {Penninx}, W. 1994, \apj, 435, 362
  
\bibitem[{{Wagoner}(1975)}]{gravrad}
{Wagoner}, R.~V. 1975, \apjl, 196, L63

\bibitem[{{White} {et~al.}(1988){White}, {Stella}, \& {Parmar}}]{spec2}
{White}, N.~E., {Stella}, L., \& {Parmar}, A.~N. 1988, \apj, 324, 363

\bibitem[{{Wijnands} \& {van der Klis}(1998)}]{source1}
{Wijnands}, R. \& {van der Klis}, M. 1998, \nat, 394, 344

\bibitem[{{Wijnands} {et~al.}(1999){Wijnands}, {van der Klis}, \&
  {Rijkhorst}}]{wvr99}
{Wijnands}, R., {van der Klis}, M., \& {Rijkhorst}, E. 1999, \apjl, 512, L39  

\bibitem[{{Wood} {et~al.}(1988){Wood}, {Ftaclas}, \& {Kearney}}]{lensing1}
{Wood}, K.~S., {Ftaclas}, C., \& {Kearney}, M. 1988, \apjl, 324, L63

\bibitem[{{Wood} {et~al.}(1991){Wood}, {Hertz}, {Norris}, {Vaughan},
  {Michelson}, {Mitsuda}, {Lewin}, {van Paradijs}, {Penninx}, \& {van der
  Klis}}]{whn+91x}
{Wood}, K.~S., {Hertz}, P., {Norris}, J.~P., {Vaughan}, B.~A., {Michelson},
  P.~F., {Mitsuda}, K., {Lewin}, W. H.~G., {van Paradijs}, J., {Penninx}, W.,
  \& {van der Klis}, M. 1991, \apj, 379, 295
  
\bibitem[{Zahn(1977)}]{zah77}
Zahn, J.-P. 1977, A\&A, 57, 383

\bibitem[{{Zhang} {et~al.}(1998){Zhang}, {Smale}, {Strohmayer}, \&
  {Swank}}]{zss+98}
{Zhang}, W., {Smale}, A.~P., {Strohmayer}, T.~E., \& {Swank}, J.~H. 1998,
  \apjl, 500, L171+

\end{thebibliography}

%% --------------------------------------------------------
%% -----------------%######################%---------------
%% -----------------%FIGURE ONE: SPECTRALS%---------------
%% -----------------%######################%---------------
%% --------------------------------------------------------
\clearpage
%% --------------------------------------------------------
%% -----------------%###############%----------------------
%% -----------------%FIGURE 2 : COMB%----------------------
%% -----------------%###############%----------------------
%% --------------------------------------------------------
\clearpage
%% --------------------------------------------------------
%% ---------------%##################%---------------------
%% ---------------%FIGURE 3 : SQUARE %---------------------
%% ---------------%##################%---------------------
%% --------------------------------------------------------
\clearpage
%% --------------------------------------------------------
%% -----------------%#######################%--------------
%% -----------------%TABLE 1 : THE BIG TABLE%--------------
%% -----------------%#######################%--------------
%% --------------------------------------------------------
\clearpage
%%%% emulateapj commands >>>>> \begin{deluxetable*}{llllcccccc}
%%%%                     >>>>> %(no tabletypesize specified)
%%%%                     >>>>> %(no tablewidth specified)
%%%%                     >>>>> %(tables 1 and 2 fit on the same page)
%%%%                     >>>>> %(no rotate)
%%%% manuscript commands >>>>> \begin{deluxetable}{llllcccccc}
%%%%                     >>>>> \tabletypesize{\tiny}
%%%%                     >>>>> %(chosen from \small |\footnotesize |\scriptsize |\tiny)
%%%%                     >>>>> \tablewidth{461.85112pt}
\begin{deluxetable}{llllcccccc}
\tabletypesize{\tiny}
\tablewidth{461.85112pt}
%% --------------------------------------------------------
\tablecaption{  
{\em{RXTE}} Observations\tablenotemark{a} of 4U~1820$-$30
\label{table1}
}
\tablehead
{
\colhead{$\;$}&
\colhead{ObsID}&
\colhead{Year}&
\colhead{Start Time\tablenotemark{b}}&
\colhead{Duration\tablenotemark{c}}&
\colhead{Time on}&
\colhead{Count Rate\tablenotemark{d}}&
\colhead{Num}&
\colhead{Num~Events\tablenotemark{f}}&
\colhead{Luminosity\tablenotemark{g}}
\\
& 
& 
& 
\colhead{(MJD)}& 
\colhead{(ks)}& 
\colhead{Source (ks)}&
\colhead{(counts/s/PCU)}& 
{PCUs\tablenotemark{e}}&
\colhead{($\times\,$10$^6$)}& 
\colhead{($\times\,$10$^{37}$~ergs/s)}
\tablecolumns{10}
}
\startdata
1 & 10075-01-01-020 &
1996 & 50384.63103 & 28.0 & 18.2 & 450 & 5 & 38.56 & 2.4 \\
2\tablenotemark{*} & 10075-01-01-02 &
1996 & 50384.96433 & 17.2 & 10.4 & 460 & 5 & 26.08 & 2.5 \\
3 & 10075-01-01-030 &
1996 & 50386.49418 & 28.2 & 17.1 & 600 -- 565 & 5 & 44.99 & 3.0 \\
4\tablenotemark{*} & 10075-01-01-031 &
1996 & 50386.82749 & 25.2 & 16.4 & 565 -- 455 & 4 & 42.17 & 2.7 \\
5 & 10075-01-01-03 &
1996 & 50387.14749 & 3.1 & 3.1 & 480 & 4 & 7.50 & 2.6 \\
6 & 20075-01-01-00 &
1997 & 50488.42272 & 24.9 & 16.0 & 635 -- 550 & 5 & 48.30 & 3.3 \\
7 & 20075-01-02-01 &
1997 & 50514.02219 & 17.9 & 11.1 & 725 & 5 & 42.19 & 3.8 \\
8 & 20075-01-03-00 &
1997 & 50530.70491 & 20.0 & 13.1 & 585 & 3 & 33.96 & 3.0 \\
9\tablenotemark{*} & 20075-01-04-00 &
1997 & 50548.70709 & 24.1 & 14.9 & 450 & 5 & 33.98 & 2.4 \\
10\tablenotemark{*} &  20075-01-06-00 &
1997 & 50595.33191 & 15.3 & 11.3 & 720 & 5 & 41.23 & 3.9 \\
11\tablenotemark{*} & 20075-01-08-00 &
1997 & 50645.22308 & 15.2 & 11.0 & 720 & 5 & 40.94 & 3.8 \\
12\tablenotemark{*} & 20075-01-09-00 &
1997 & 50675.45754 & 25.1 & 16.0 & 670 -- 735 & 4 & 51.31 & 3.6 \\
13 & 30057-01-01-02 &
1998 & 50909.66291 & 18.4 & 11.2 & 470 & 4 & 24.24 & 2.5 \\
14\tablenotemark{*} & 30053-03-01-000 &
1998 & 50920.19895 & 25.2 & 16.4 & 465 & 5 & 39.94 & 2.5 \\
15 & 30053-03-01-001 &
1998 & 50920.52779 & 20.8 & 12.0 & 465 & 4 & 24.94 & 2.5 \\
16\tablenotemark{*} & 30053-03-01-00 &
1998 & 50920.79458 & 25.8 & 13.2 & 475 & 5 & 41.05 & 2.6 \\
17 & 30053-03-02-03 &
1998 & 51080.45284 & 20.9 & 14.2 & 430 -- 380 & 4 & 22.30 & 2.2 \\
18 & 40017-01-02-00 &
1999 & 51222.63887 & 20.9 & 14.4 & 510 & 4 & 33.63 & 2.9 \\
19 & 40017-01-03-00 &
1999 & 51238.03748 & 15.1 & 10.4 & 320 & 5 & 16.81 & 1.8 \\
20 & 40017-01-04-00 &
1999 & 51253.69624 & 18.9 & 10.5 & 430 & 4 & 13.42 & 2.4 \\
21\tablenotemark{*} & 40017-01-07-00
& 1999 & 51300.07107 & 18.6 & 10.9 & 980 -- 895 & 3 & 37.42 & 5.0 \\
22 & 40017-01-09-00 &
1999 & 51330.51073 & 15.1 & 10.6 & 700 & 3 & 28.80 & 3.9 \\
23 & 40017-01-12-00 &
1999 & 51389.38952 & 15.3 & 10.7 & 705 & 3 & 23.53 & 3.8 \\
24\tablenotemark{*} & 40017-01-13-00 &
1999 & 51400.44912 & 21.0 & 14.7 & 660 & 3 & 28.88 & 3.6 \\
25 & 40017-01-14-00 &
1999 & 51407.50805 & 15.1 & 10.6 & 505 & 5 & 27.13 & 2.9 \\
26 & 30057-01-03-01 &
1999 & 51410.50430 & 24.0 & 15.1 & 515 -- 445 & 3 & 24.59 & 2.5 \\
27 & 30057-01-03-000 &
1999 & 51412.31866 & 29.0 & 16.9 & 495 -- 530 & 3 & 34.10 & 2.9  \\
28 & 30057-01-03-00 &
1999 & 51412.65197 & 7.5 & 5.1 & 510 & 4 & 10.74 & 2.7 \\
29 & 40017-01-15-00 &
1999 & 51417.29843 & 20.4 & 12.8 & 440 & 3 & 17.05 & 2.4 \\
30 &  40019-02-01-00 &
1999 & 51421.30667 & 25.0 & 15.3 & 370 & 3 & 16.81 & 2.0 \\
31 & 30057-01-05-00 &
1999 & 51435.00244 & 26.8 & 17.9 & 365 -- 480 & 2 & 26.34 & 2.4 \\
32 & 40017-01-18-00 &
1999 & 51464.36730 & 20.8 & 14.0 & 535 & 3 & 26.07 & 3.0 \\
33 & 40017-01-21-00 &
2001 & 51941.01232 & 14.9 & 10.4 & 380 & 4 & 14.36 & 2.0 \\
34\tablenotemark{*} &
40017-01-23-01 &
2002 & 52355.88359 & 14.8 & 10.3 & 915 & 3 & 32.89 & 4.8 \\
\enddata
\tablenotetext{a}{We used event data with $\sim$125~$\mu$s time resolution
                  for all observations.}
\tablenotetext{b}{Start time of the observation in MJD.}
\tablenotetext{c}{Stop time $-$ Start time of the observation in ks.}
\tablenotetext{d}{Approximate photon count per second per PCU obtained
                  from the Standard~2 lightcurve in the wide band (2-20~keV).}
\tablenotetext{e}{Minimum number of PCUs in use at a given
                  moment of the observation.}
\tablenotetext{f}{Approximate total number of events
                  processed in the wide band (2--20~keV).}
\tablenotetext{g}{2--10~keV luminosities obtained from spectral fitting
                  and based on an assumed distance of 6.4~kpc.}
\tablenotetext{*}{Observations marked with a $^{*}$ were
                  searched with acceleration searches, phase modulation searches,
		  and an orbital-parameter-fitting coherent search. The remaining
		  observations were searched only with acceleration
		  searches and phase modulation searches.}
\end{deluxetable}
%% --------------------------------------------------------
%% -----------------%########################%-------------
%% -----------------%TABLE 2 : THE WIDE TABLE%-------------
%% -----------------%########################%-------------
%% --------------------------------------------------------
\clearpage
%%%% emulateapj commands >>>>> \begin{deluxetable*}{lll}
%%%%                     >>>>> %(no tabletypesize specified)
%%%%                     >>>>> %(no tablewidth specified)
%%%%                     >>>>> %(tables 1 and 2 fit on the same page)
%%%%                     >>>>> %(no rotate)
%%%% manuscript commands >>>>> \begin{deluxetable}{lll}
%%%%                     >>>>> \tabletypesize{\tiny}
%%%%                     >>>>> %(chosen from \small |\footnotesize |\scriptsize |\tiny)
%%%%                     >>>>> \tablewidth{443pt}
\begin{deluxetable}{ll}
\tabletypesize{\tiny} 
\tablewidth{443pt}
%% --------------------------------------------------------
\tablecaption{
Search Types for Different Data File Types
\label{table2}
}
\tablehead{
\colhead{Type of Data File} & 
\colhead{Searches Run}
}
\startdata
Observation-long lists of events &
orbital-parameter-fitting coherent search\tablenotemark{a} (W band only)\\
Observation-long time series binned at 0.244~ms &
Acceleration searches (all bands),
phase modulation searches (all bands)\\
{\em{RXTE}}-orbit-long lists of events & No searches run\\
{\em{RXTE}}-orbit-long time series
binned at 0.122~ms &
Acceleration searches (all bands),
phase modulation searches (all bands)\\
600-s-long and 1500-s-long lists of events & No searches run\\
600-s-long and 1500-s-long time series
binned at 0.122~ms &
Acceleration searches (all bands)\tablenotemark{b}\\
\enddata
\tablenotetext{a}
{
Because running an orbital-parameter-fitting coherent search was extremely
demanding computationally, we ran this search on event lists of 11
observations only. By contrast, we ran accelerations searches and phase
modulation searches on time series of all 34 observations.
}
\tablenotetext{b}
{
The segmented acceleration searches that we ran, even though they were not
short enough for the acceleration to be constant, could still be useful if a
strong pulsed signal were to appear for a short period of time.
}
\end{deluxetable}
%% --------------------------------------------------------
%% ---------------%########################%---------------
%% ---------------%TABLE 3 : WITH DELUX    %---------------
%% ---------------%########################%---------------
%% --------------------------------------------------------
\clearpage
%%%% emulateapj commands >>>>> \begin{deluxetable}{ccc}
%%%%                     >>>>> %(no tablewidth specified)
%%%% manuscript commands >>>>> \begin{deluxetable}{ccc}
%%%%                     >>>>> \tablewidth{249.37006pt}
\begin{deluxetable}{ccc}
\tablewidth{249.37006pt}
%% --------------------------------------------------------
\tablecaption
{
Results of Sensitivity Calculations for the
Orbital-Parameter-Fitting Coherent Searches
\label{table3}
}
\tablehead
{
\colhead{Pulsed Fraction} & 
\colhead{Best Case} & 
\colhead{Worst Case} \\
& 
\colhead{Detection $\sigma$\tablenotemark{*}} & 
\colhead{Detection $\sigma$\tablenotemark{*}}
}
\startdata
0.17\% & 3 & $<$~3 \\
0.2\% & 5.3 & $<$~3 \\
0.26\% & $>$~5.3 & 3 \\
0.3\% & $>$~5.3 & 5.3 \\
\enddata
\tablenotetext{*}
{
$\sigma$ is an equivalent Gaussian significance.
}
\end{deluxetable}
\clearpage
\end{document}